\documentclass[pra,twocolumn,showpacs]{revtex4}

\usepackage{amssymb,graphicx}
\usepackage{dsfont}
\usepackage{pdflscape}
\usepackage{rotating}
\usepackage{array}
\usepackage{wasysym}

\begin{document}

\title{Distillation of mixed-state continuous-variable entanglement by photon subtraction}
\author{ShengLi Zhang and Peter van Loock}\email{peter.vanloock@mpl.mpg.de}

\affiliation{Optical Quantum Information Theory Group, Max Planck Institute for the Science of Light, G\"unther-Scharowsky-Str.1/Bau 26, 91058 Erlangen, Germany}
\affiliation{Institute of Theoretical Physics I, Universit\"at Erlangen-N\"urnberg, Staudstr.7/B2, 91058 Erlangen, Germany}


\begin{abstract}
We present a detailed theoretical analysis for the distillation of one copy
of a mixed two-mode continuous-variable entangled
state using beamsplitters and coherent
photon-detection techniques, including conventional on-off detectors
and photon number resolving detectors.
The initial Gaussian mixed-entangled states are generated by transmitting
a two-mode squeezed state through a lossy bosonic channel, corresponding to the primary
source of errors in current approaches to optical quantum communication.
We provide explicit formulas to calculate the entanglement
in terms of logarithmic negativity before and after distillation, including losses in
the channel and the photon detection, and show that
one-copy distillation is still possible even for losses near the typical
fiber channel attenuation length. A lower bound for the
transmission coefficient of the photon-subtraction beamsplitter is derived, representing
the minimal value that still allows to enhance the entanglement.

\end{abstract}
\maketitle


 \section{Introduction}

Entanglement of composite systems represented by continuous quantum variables
is of conceptual importance for studying
fundamental questions of quantum mechanics \cite{EPR} and promises to be
useful for potential real-world applications
in the fast developing field of quantum information
processing \cite{RMP,eisert03,adesso07}.
However, in general, entanglement is a fragile resource which is easily degraded
when it interacts with an uncontrollable environment, for example,
in a communication channel in form of a lossy and noisy, optical fiber.
In order to overcome this problem of entanglement degradation, typically,
protocols such as entanglement purification or
distillation will be utilized, as originally proposed for qubits \cite{bennet}.
These schemes detect errors and are usually probabilistic,
as opposed to deterministic approaches such as quantum error correction.
More generally, entanglement distillation can be defined as any scheme that
creates one or more entangled pairs of higher entanglement from one or more copies
of initially imperfectly entangled pairs by means of local operations
and classical communication.

Although many impressive experiments for the
distillation of pure or mixed discrete-variable
entanglement have been reported \cite{Kwiat01,Pan01,Yamamoto03,Reichle06},
distilling continuous-variable entanglement appears to be
rather different and, in general, harder to achieve.
The difficulty arises mainly from the necessity
of a non-Gaussian element for distilling Gaussian entangled states \cite{nogo1,nogo2,nogo3}.
For instance, in order to distill the quantum optical two-mode squeezed state (TMSS)
whose quadratures obey Gaussian statistics, one must introduce at least
one non-Gaussian operation, in form of a non-Gaussian ancilla or a non-Gaussian
measurement. The so-called photon
subtraction (PS) strategy, first introduced by Opatrn\'{y} \emph{et al.} \cite{Opatrny},
is one of the experimentally most readily available
operations beyond the Gaussian regime.
It enables one to modify the Gaussian statistics of a given TMSS and therefore serves
as a possible approach to entanglement distillation of such Gaussian states
\cite{MSKIM}.
The basic principle of the PS technique is very simple and can be
implemented using a beamsplitter and photon measurements.

After Opatrn\'{y} \emph{et al.'s} pioneering work \cite{Opatrny}, many
efforts have been made to improve the performance of such an entanglement distillation protocol.
Olivares \emph{et al.} \cite{Oliver} proposed an inconclusive
PS method, which employs a more realistic on-off photon detector in order to enhance the entanglement.
Kitagawa \emph{et al.} \cite{Kitagawa} presented a fairly complete theoretical
analysis of this type of distillation, including
a numerical evaluation of the entanglement before and after distillation.
Moreover, a multi-mode theory for frequency mode matching in the photon-subtracting operation
was derived \cite{ModMatching}.
Even an experiment implementing Opatrn\'{y}'s method has already been reported \cite{NatPhotonic}.
These efforts, both on the theoretical and the
experimental side, are examples for the more recent attempts to combine
discrete-variable and continuous-variable approaches to optical
quantum information processing \cite{pvllaserphreviews}.

The original scheme by Opatrn\'{y} \emph{et al.} as well as its
theoretical refinements and extensions all refer to a single copy
of a {\it pure}, Gaussian entangled state which is distilled into a non-Gaussian
entangled state. This kind of distillation is sometimes referred to
as entanglement concentration, distinct from entanglement purification protocols
in which initially mixed-entangled states are purified and thereby turned
into states with higher entanglement. Even though usually such entanglement purification
is applied to two or more copies of entangled states \cite{bennet},
one mixed-entangled copy may also be distilled through local, generalized measurements,
similar to those for concentrating a single pure-state copy into a maximally
entangled state \cite{gisin,kent,verstraete,wangPRL}. In the mixed-state case,
however, both parties sharing the initial state must perform local measurements
and communicate their results.

In this paper, we provide a detailed analysis for the one-copy distillation of
\emph{mixed} continuous-variable entanglement,
using beamsplitters and experimentally feasible photon detection techniques.
In other words, similar to those one-copy schemes
mentioned in the preceding paragraph, we shall consider non-Gaussian,
generalized measurements locally performed on the two modes of the initial Gaussian state.
Note that PS ideally corresponds to maps like
$\hat a |n\rangle = \sqrt{n}|n-1\rangle$ or
$\hat a |\alpha\rangle = \alpha|\alpha\rangle$, with the photon annihilation
operator being non-unitary and the resulting states being not normalized
-- a manifestation of the probabilistic nature of the PS process.
The corresponding generalized measurements are realized through beam splitter
transformations, locally acting on the two
signal modes and additional ancilla vacuum modes,
and subsequent photon measurements. Note that various experiments
have already demonstrated how versatile PS is for non-Gaussian state engineering
\cite{wenger,ourjoumtesv,parigi}.
Local filters for entanglement concentration of one copy of a pure TMSS
were previously considered using
Kerr interactions or CQED \cite{fiurasekandmista}.
Multi-copy distillations of noisy versions of a TMSS have been proposed as well
\cite{browneandeisert,lundandralph,duanQND}.

A further recent one-copy scheme for distilling mixed continuous variable
entanglement was proposed in Ref.~\cite{Rebic}. Different from our analysis,
this proposal \cite{Rebic} employs detections of collective
excitations in atomic ensembles for the non-Gaussian operations. Further, it
uses only operational entanglement measures, namely teleportation fidelities.
Here, we shall calculate both fidelities and, in particular,
logarithmic negativities for the distilled states.

While in the theoretical analysis of Ref.~\cite{Kitagawa},
the initial states are pure (and become mixed only for the case of on-off detections),
the input states of the recent experiment \cite{NatPhotonic} were mixed
due to experimental imperfections such as the complication to prepare
a perfectly pure, minimum-uncertainty squeezed state.
Although even the idealized pure-state versions of that experiment
slightly differ from the TMSSs used in our analysis
(by local squeezing operations), the present article also provides
a more general theoretical foundation of the experiment
described in Ref.~\cite{NatPhotonic}. It proves
the possibility and feasibility of realistic PS-based distillation
of a TMSS, even in the presence of high losses. At the same time it illuminates
the applicability of Opatrn\'{y}'s PS distillation protocol \cite{Opatrny}
and provides more details on how to improve entanglement in a general,
realistic mixed-state scenario; for instance, in optical quantum communication using
lossy fibers of nearly one attenuation length as potentially used in
a quantum repeater \cite{duerandbriegel}. Moreover, some of our results
are fully analytical, thus further extending the theory presented in Ref.~\cite{Kitagawa}.

The paper is organized as follows. Firstly, in Sec.~II, we will give a description of our scheme for
entanglement distillation, along with the method for generating of our mixed entangled state.
In Sec.~III, the definition of logarithmic negativity is briefly summarized.
With such a figure of merit for entanglement, the amount of entanglement before
distillation is explicitly derived. Sec.~IV is devoted to the
entanglement distillation with conventional on-off detectors.
An exact analytical formula for the entanglement distillation of pure TMSSs
is obtained, which previously was only numerically investigated by
Kitagawa \textit{et al.} \cite{Kitagawa}. In Sec.~V, we discuss two different
strategies of applying photon number resolving detectors (pure and mixed PNR detectors).
The success probability of distillation
and the corresponding lower bounds $T_L$ are studied, respectively.
In Sec.~\ref{fidelitySec}, to further illustrate our results,
we calculate an operational
measure of entanglement (the fidelity in quantum teleportation), leading to
yet another way to compare the entangled states before and after
distillation. Finally, we conclude with a discussion and summary in Sec.~VII.

\section{Lossy bosonic channel and photon subtraction}

Let us first introduce the amplitude-damped TMSS with which we are mainly concerned
in this paper. This kind of mixed-state entanglement can be obtained by
sending each mode of a pure TMSS through a lossy bosonic channel.
In our theoretical analysis, we shall simulate the lossy channel through an
extra vacuum mode and a beam splitter.
This is the simplest model for mimicking realistic fiber-optical light propagation,
where more and more signal photons are gradually absorbed on the way during the channel
transmission.

\begin{figure}
  \includegraphics[width=8.4 cm]{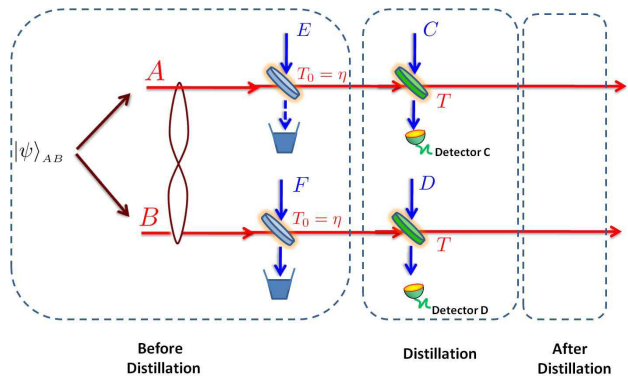}\\
  \caption{Scheme of continuous-variable entanglement distillation.
 The initial state $|\psi\rangle_{AB}$ is a pure two-mode squeezed vacuum state,
both beams of which are transmitted through beamsplitters
 with transmittance $T_0=\eta$ in order to simulate
 a lossy bosonic channel of transmission $\eta$.
The beam splitters with transmittance $T$ including the photon detectors
are used to achieve the photon subtractions for
entanglement distillation. The input states of the $C,D,E,F$-modes
are pure vacuum states. An event of successful distillation is heralded when both
detectors register non-zero counts.
  }\label{SchemefigONOFF}
\end{figure}

The entire scheme for our distillation protocol is shown in
Fig.~\ref{SchemefigONOFF}. The initial pure TMSS is given by
\begin{eqnarray}
|\psi\rangle_{AB} &=&\sum\limits_{n=0}^\infty \alpha_n|n\rangle_A |n\rangle_B ,~~~\alpha_n=\sqrt{1-\lambda^2}\lambda^n,
\end{eqnarray}
with $\lambda = \tanh(r)$ representing the degree of squeezing and $A,B$ referring to
the two transmitted modes. Two beamsplitters with transmission coefficient $T_0=\eta$
and auxiliary vacuum modes $E, F$ are put into the ideal channels
in order to simulate two lossy channels with
transmission efficiency $\eta$. The PS-based distillation is implemented
using two more beam splitters (each with transmission $T$) and photon detectors. Due to
the finite reflectance of these beam splitter, the photon subtraction process
heralded by the photon
detectors is a probabilistic process and successful distillation may occur whenever
both detectors register non-zero counts.

To give a systematical analysis, in this paper, we shall consider entanglement
distillation with two different types of detectors, namely
conventional on-off detectors (e.g., avalanche
photondiodes, APD) and photon number resolving
detectors (PNR), both of which are widely used in quantum optical
experiments.

\section{Logarithmic Negativity and entanglement Before Distillation}
\label{SecBFdis}

Following the definitions of Ref.~\cite{Kitagawa}, we will use the
logarithmic negativity as a figure of merit to quantify entanglement. The
logarithmic negativity \cite{LN,Plenio,eisert01} is a relatively easily computable measure of entanglement;
more precisely, it is an entanglement monotone, both under
local operations and classical communication
and under positive partial transpose preserving operations.

The {\it logarithmic negativity} of a bipartite state
 $\rho_{_{AB}}=\rho$ is defined by
\begin{eqnarray}
E_N(\rho)=\log_2(1+2N(\rho)),\label{ENdef}
\end{eqnarray}
in which $N(\rho)$ is defined as the \textit{negativity}
given by
the absolute value of the sum of negative eigenvalues of the partially
transposed density operator $\rho^{\Gamma_{A}}$. Here and throughout,
without loss of generality,
we will perform the partial transpose operation with respect to the $A$-mode.

We shall now quantify the amount of entanglement of the amplitude-damped TMSS.
Note that this state including the damping effect remains a Gaussian state,
and hence its logarithmic negativity could be directly computed from
its second-moment covariance matrix through the corresponding symplectic eigenvalues
of the partially transposed state \cite{RMP,eisert03,adesso07}. However, for our purposes,
since the PS process will lead to non-Gaussian states, it is more useful
to achieve a more general entanglement quantification expressed in the photon number basis.
This is similar to the approach of Ref.~\cite{Kitagawa}, but
with the important distinction that our states are mixed states from the beginning,
both before and after the distillation.

First of all, let us denote the beamsplitter coupling between modes $k$ and $l$
as \cite{Kitagawa},
\begin{eqnarray}
V_{kl}(\theta_0)=\exp\left[\theta_0(a_{k}^\dagger a_{l}-a_{k} a_{l}^\dagger)\right],
\end{eqnarray}
with $\theta_0=\arctan(\sqrt{(1-\eta)/\eta})$ and $a_{k(\l)}$ being the photon
annihilation operator of the $k(l)$-mode. The unitary state evolution before entanglement
distillation can be formulated as follows,
\begin{eqnarray}
|\Psi\rangle_{_{ABEF}}=V_{_{AE}}(\theta_0)\otimes V_{_{BF}}(\theta_0)|\psi\rangle_{_{AB}}|0\rangle_{_E}|0\rangle_{_F},\label{Eq1}
\end{eqnarray}
where $|0\rangle_{_E}, |0\rangle_{_F}$ are the initial vacuum states
of the loss modes.
Through direct calculation, we have
\begin{eqnarray}
&&|\Psi\rangle_{_{ABEF}}=\sum\limits_{n=0}^{\infty}\sum\limits_{k,l=0}^{n}\alpha_n\xi_{nk}\xi_{nl}|n-k\rangle_{_A}|n-l\rangle_{_B}|k\rangle_{_E}|l\rangle_{_F},\nonumber\\
&&\xi_{nm}=(-1)^m\sqrt{{n \choose m}}(\sqrt{\eta})^{n-m}(\sqrt{1-\eta})^m,
\label{xiDef}
\end{eqnarray}
where $m=0,1,\cdots,n$ and ${n \choose m}=\frac{n!}{m!(n-m)!}$ is
the binomial coefficient.

The mixed state $\rho_{_{AB}}$ before entanglement distillation is obtained
 by tracing over the loss modes $E$ and $F$,
\begin{eqnarray}
\rho_{mix}&&\equiv\rho_{_{AB}}=\mathrm{Tr}_{_{EF}}[|\Psi\rangle_{_{ABEF}}\langle\Psi|]\nonumber
\\&&=\sum\limits_{m,n=0}^{\infty}\sum\limits_{i=0}^{n}\sum\limits_{j=0}^{m}f_{nmij}|n-i\rangle_{_A}\langle m-i|\otimes\nonumber\\
&& ~~~~~~~~|n-j\rangle_{_{B}}\langle m-j|,\label{rhoABbeforeDist}
\end{eqnarray}
with $f_{nmij}$ being a real positive coefficient, $f_{nmij}=\alpha_n\alpha_m\xi_{ni}\xi_{mi}\xi_{nj}\xi_{mj}$.

Similar to the case of a pure TMSS \cite{Kitagawa,PRAHK}, the partial transpose of the density
 matrix (\ref{rhoABbeforeDist}) is still block diagonal in the photon number (Fock) basis. We have
\begin{eqnarray}
&&\rho_{_{AB}}^{\Gamma_A}=\sum\limits_{m,n=0}^{\infty}\sum\limits_{i=0}^{n}\sum\limits_{j=0}^{m}f_{nmij}|m-i,n-j\rangle_{_{AB}}\langle n-i,m-j|\nonumber\\
&&=\bigoplus_{K=0}^\infty\sum\limits_{i,j=0}^{K} C_{i,j}^{(K)}|i,K-i\rangle_{_{AB}}\langle j, K-j|,\label{CK}
\end{eqnarray}
 with
\begin{eqnarray}
C_{i,j}^{(K)}&&=(1-\lambda^2)\left(\frac{\eta}{1-\eta}\right)^K\sum\limits_{n=n_0}^\infty\sqrt{{K\choose i}{K\choose j}}\nonumber\\
&&~~~~\times (\lambda-\lambda\eta)^{(i+j+2n)}\frac{(n+i)!(n+j)!}{K!n!(n+i+j-K)!},\nonumber\\
n_0&&=\max\{0,K-i-j\}.\label{CKdef}
\end{eqnarray}

Thus, the negativity of $\rho_{mix}$ can be equivalently obtained by solving
the eigenvalue problem of a series of $(K+1)\times(K+1)$ sub-matrices
\begin{eqnarray}
\mathbf{C}_\mathrm{K}=\left[C_{i,j}^{(K)}\right]_{i=0,\cdots,K;~j=0,\cdots,K},\label{Ckbefore}
\end{eqnarray}
for $K=0,1,\cdots,\infty$.
Such a block sub-matrix method is quite efficient in numerical
 evaluations \cite{Kitagawa}.
Indeed, the matrix $\mathbf{C}_\mathrm{K}$ has a very useful symmetry property
which will finally simplify the whole problem.

\noindent\textbf{Theorem 1.} $\mathbf{C}_\mathrm{K}$ is a double
symmetric, i.e., both symmetric and centrosymmetric matrix.

\noindent\textit{Proof.} Symmetric property follows directly from the
$i,j$ exchange invariance in the definition Eq.~(\ref{CKdef}).
Therefore we only need to prove $C_{i,j}^{(K)}=C_{K-i,K-j}^{(K)}$ for arbitrary $i,j$.
Now consider any $i,j$ such that for $i+j\le K$, we always have $(K-i)+(K-j)\ge K$.
Using Eq.~(\ref{CKdef}), it directly follows $n_0(i,j)=K-i-j\ge 0$ and $n_0(K-i,K-j)=0$.
By replacing the index $n=n^{\prime}+n_0(i,j)$ in the summation of $C_{i,j}^{(K)}$ and
noticing ${K \choose i}={K\choose K-i}$, the relation $C_{i,j}^{(K)}=C_{K-i,K-j}^{(K)}$
can be straightforwardly obtained.

\noindent \textbf{Theorem 2.}
The negativity of $\rho_{mix}$ can
 be uniquely determined by the skew diagonal entries of matrix  $\mathbf{C}_\mathrm{K}$:
\begin{eqnarray}
N(\rho_{mix})=\frac{1}{2}\left(\sum_{K=0}^\infty\mathrm{Tr}[\mathbf{J}_\mathrm{K} \mathbf{C}_\mathrm{K}]-1\right),
\end{eqnarray}
with $\mathbf{J}_\mathrm{K}$ being the anti-identity matrix,
\begin{eqnarray}
\mathbf{J}_\mathrm{K}=\left[\delta_{i+j,K}\right]_{i,j=0,\cdots,K}=\left(
\begin{array}{ccccc}
0 & 0 & \cdots & 0  & 1\\
0 & 0 & \cdots & 1  & 0 \\
\vdots & \vdots & \ddots &  \vdots & \vdots\\
0 & 1 & \cdots &  0 & 0\\
1 & 0 &\cdots &0 &0
\end{array}%
\right).
\end{eqnarray}
\noindent\textit{Proof}: According to Ref.~\cite{CentroSym}, for
a $(K+1)\times(K+1)$ dimensional matrix $\mathbf{C}_{\mathrm{K}}$,
 there always exists an orthogonal matrix $U$, such that for $K$ odd,
\begin{eqnarray}
U\mathbf{C}_{\mathrm{K}}U^T=
\left(
\begin{array}{cc}
A-J M & 0\\
0  &  A+J M
\end{array}%
\right),
\end{eqnarray}
and for $K$ even,
\begin{eqnarray}
U\mathbf{C}_{\mathrm{K}}U^T=
\left(
\begin{array}{ccc}
A-J M & 0 & 0\\
0  &  q& \sqrt{2}x^T\\
0 & \sqrt{2} x& A+J M  \end{array}%
\right),\label{AJC}
\end{eqnarray}
where $A$, $M$, and $J$ are each $\lfloor\frac{K+1}{2}\rfloor\times\lfloor
\frac{K+1}{2}\rfloor$ matrices with elements
\begin{eqnarray}
A_{i,j}=C^{(K)}_{i,j},M_{i,j}=C^{(K)}_{ i+ \lceil\frac{K-1}{2}\rceil,j},
~J_{i,j}=\delta_{i+j,\lfloor\frac{K-1}{2}\rfloor},\nonumber
\end{eqnarray}
for $i,j=0,1,\cdots,\lfloor (K-1)/2\rfloor$, and $\lfloor\cdots\rfloor$ and
$\lceil\cdots\rceil$ are the floor and ceiling functions, respectively.

This shows that the eigenvalues of $\mathbf{C}_{\mathrm{K}}$ are the same as
 the eigenvalues of $A-JM$ and $A+JM$ in the case of $K$ odd and the same as $A-JM$ and
$\left(\begin{array}{cc}
q & \sqrt{2}x^T\\
\sqrt{2}x  &  A+JM
\end{array}\right)$ in the case of $K$ even.

Moreover, it can be always checked that the sub-block $A-JM$ contains all the negative
 eigenvalues of matrix $\mathbf{C}_{\mathrm{K}}$. In fact, for the matrix
 $\mathbf{C}_{\mathrm{K}}$ as defined in Eq.(\ref{CKdef}),
 $A-JM$ is always negative-definite, whereas the sub-block $A+JM$ and  $
\left(\begin{array}{cc}
q & \sqrt{2}x^T\\
\sqrt{2}x  &  A+J M
\end{array}\right)$ are positive-definite. Thus,
the absolute value of the sum of the negative eigenvalues
 of $\mathbf{C}_{\mathrm{K}}$ is given by
$|\mathrm{Tr}[A-JM]|=\mathrm{Tr}[JM-A]=\frac{1}{2}\mathrm{Tr}[
\mathbf{J}_\mathrm{K}\mathbf{C}_\mathrm{K}-\mathbf{C}_\mathrm{K}]
$.
The negativity of the whole matrix $\rho_{mix}$ follows as
\begin{eqnarray}
N(\rho_{mix})&=&\frac{1}{2}\sum_{K=0}^\infty\mathrm{Tr}[\mathbf{J}_\mathrm{K}\mathbf{C}_\mathrm{K}-\mathbf{C}_\mathrm{K}]\nonumber\\
&=&\frac{1}{2}\left(\sum_{K=0}^\infty\mathrm{Tr}[\mathbf{J}_\mathrm{K} \mathbf{C}_\mathrm{K}]-1\right),
\end{eqnarray}
where in the second line, we have imposed the normalization condition
 \begin{eqnarray}
\sum_{K=0}^\infty\mathrm{Tr}[\mathbf{C}_{\mathrm{K}}]
=\mathrm{Tr}(\rho_{_{AB}}^{\Gamma_{A}})=\mathrm{Tr}(\rho_{_{AB}})=1.
\end{eqnarray}

Thus, following the definition in Eq.~(\ref{ENdef}), the logarithmic
negativity of the state in Eq.~(\ref{rhoABbeforeDist}) can now be easily calculated as
\begin{eqnarray}
E_{N}(\rho_{mix})&=&\log_2\left(\sum_{K=0}^\infty \mathrm{Tr}[\mathbf{J}_\mathrm{K} \mathbf{C}_\mathrm{K}]\right)\nonumber\\
&=&\log_2\frac{1+\lambda}{1-\lambda(2\eta-1)}. \label{EnrhoBeforeDis}
\end{eqnarray}
By setting $\eta=\exp(-\gamma t)$, our result agrees with that presented in
 Ref.~\cite{PRAHK}; however, our derivation leads to a simple, closed expression
 as a function of the input squeezing and channel loss. For this still Gaussian state
 before distillation, we can also confirm our result by calculating the
 symplectic eigenvalues of the partially transposed state.

Following our formalism above, we merely need to calculate the skew diagonal
entries $\{C_{i,K-i}^{(K)}\}_{i=0,\cdots,K}$ in order to obtain the logarithmic
negativity for the two-mode mixed entangled state. It is important to
note that Theorem 1 and Theorem 2 can also be applied to
calculate the logarithmic negativity after entanglement
distillation. In fact, PS on both transmitted modes
does not change the symmetry and centrosymmetry of the partially
transposed density matrix $\rho_{_{AB}}^{\Gamma_{_A}}$, provided that {\it both
detectors obtain the same measurement results}. The only difference is that the state after
PS is not normalized. One should then specify the normalization factor
(i.e., the trace of $\rho_{_{AB}}^{dist}=\sum_K\mathrm{Tr}\left[\mathbf{C}_\mathrm{K}\right]$)
for the different types of detectors and the different detection strategies.
This enables us to extend the analytical formulas for
continuous-variable entanglement
from pure states to mixed states, including the Gaussian state
$\rho_{mix}$ before distillation as well as
the non-Gaussian states after distillation using on-off detectors or
PNR detectors (in pure and mixed
strategies, see below).

\section{Distillation using On-Off detection}\label{on-offSec}

For convenience, let us first give a general description of photon detectors.
Suppose the detector can respond with $\mathcal{M}$ different measurement outcomes.
According to the theory of generalized quantum measurements \cite{POVM1,POVM2},
such a measurement device can be
completely characterized through a set of positive-definite operators
 $\{\hat{\Pi}_{k}|{k=1,2,\cdots,\mathcal{M}}\}$, corresponding
 to a positive operator-valued measure (POVM). The quantum
 measurement is probabilistic: for a given input state $\varrho$, the probability that
the detector gives
outcome $k$ is $P_k=\mathrm{Tr}[\hat{\Pi}_{k}\varrho]$. The condition that
the total probability is normalized corresponds to $\sum_{k=1}^{\mathcal{M}}\hat{\Pi}_k=\mathds{1}$,
with $\mathds{1}$ representing the identity operator.

The photon detectors usually employed in quantum optical experiments, such as
 avalanche photodiodes (APD) operating in the Geiger mode, correspond to a measurement
device with only two measurement outcomes: off (no photons detected) and on
(one or more photons detected). Expressed in the Fock basis, the positive operator description
of an ideal on-off photon detector is then given by  $\{\hat{\Pi}^{(off)},\hat{\Pi}^{(on)}\}$, with
\begin{eqnarray}
&&\hat{\Pi}^{(off)}=|0\rangle\langle 0|,\nonumber
 \\&&\hat{\Pi}^{(on)}=\mathds{1}-\hat{\Pi}^{(off)}=\sum_{k=1}^\infty |k \rangle\langle k|.
\end{eqnarray}

Based on the formalism and the notations above, we can now proceed with the entanglement distillation
protocol
in Fig.\ref{SchemefigONOFF}. Assuming that the two beamsplitters for PS have
the same transmittance $T$ (reflectance coefficient $R=1-T$), the
state evolution of the whole PS process can be described by
\begin{eqnarray}
\rho_{_{ABCD}}&&=\mathds{V}\left[\rho_{mix}\otimes|0\rangle_{_C}\langle 0|\otimes|
0\rangle_{_D}\langle 0|\right]\mathds{V}^\dagger\label{rhoABCD},\\
\widetilde\rho(on,on)&&=\frac{\mathrm{Tr}_{_{CD}}\left[\rho_{_{ABCD}}\mathds{1}_{_{AB}}\otimes\hat{\Pi}_{_C}^{(on)}
\otimes\hat{\Pi}_{_D}^{(on)}\right]}{P(on,on)},\nonumber
\end{eqnarray}
where $\mathds{V}=V_{_{AC}}(\theta)\otimes V_{_{BD}}(\theta)$, $\theta=\sqrt{R/T}$, and
$\widetilde\rho(on,on)$ is the normalized output state; $P(on,on)$ is the
 probability of detecting non-zero photons in both detectors,
\begin{eqnarray}
P(on,on)=\mathrm{Tr}\left[\rho_{_{ABCD}}\mathds{1}_{_{AB}}\otimes\hat{\Pi}_{_C}^{(on)}\otimes\hat{\Pi}_{_D}^{(on)}\right],
\end{eqnarray}
where this time the trace is over all four modes ABCD.
Using the same method as in Sec.~\ref{SecBFdis}, analytic formulas for the entanglement and
the success probability can now be derived.
The unnormalized, partial transpose $\rho_{AB}^{\Gamma_{A}}$ is again block diagonal with respect to the $K$-subspaces.
We have
\begin{widetext}
\begin{eqnarray}
&&C_{i,j}^{(K)} (on,on)\nonumber\\&&=(1-\lambda^2)\left(\frac{\eta T}{1-\eta}\right)^K
\sum\limits_{\gamma=1}^\infty \sum\limits_{\delta=1}^\infty\sum\limits_{n=n_0}^\infty\left(\frac{\eta R}{1-\eta}\right)^{\gamma+\delta}(\lambda-\lambda\eta)^{i+j+2n+2\gamma}\frac{(i+n+\gamma)!(j+n+\gamma)!}{K!n!\gamma!\delta!(n+i+j+\gamma-K-\delta)!}\sqrt{{K\choose i}{K \choose j}},\nonumber
\\
\nonumber\\
n_0&&=\max\{0,K+\delta-i-j-\gamma\}.\label{CKdefonon}
\end{eqnarray}
\end{widetext}

The probability of success can be evaluated as
\begin{eqnarray}
P(on,on)&=&\sum_{K=0}^{\infty} \sum_{i=0}^{K} C_{i,i}^{(K)} (on,on)\nonumber\\
&=&\frac{\lambda^2 (1-\widetilde{T})^2 (1+\lambda^2 \widetilde{T})}{(1-\lambda^2 \widetilde{T})(1-\lambda^2 \widetilde{T}^2)},\label{Ponon}
\end{eqnarray}
where we define $\widetilde{T}=1-\eta R$ and $\widetilde{R}=1-\eta T$.

After state normalization, the logarithmic negativity can be also analytically obtained:
\begin{widetext}
\begin{eqnarray}
E_N(\widetilde{\rho}(\textit{on,on}))&=&
\log_2\left[\frac{(1-\lambda^2)\eta R}{(1-\lambda \eta T)^2-\lambda^2 (1-\eta) \widetilde{R}}\right]
+\log_2\left[\frac{\widetilde{R}}{(1-\lambda)(1-\lambda(2\eta T-1))}-\frac{1-\eta}{(1-\lambda \eta T)^2 -\lambda^2 (1-\eta)^2}\right]\nonumber\\
&&+\log_2\left[\frac{(1-\lambda^2 \widetilde{T})(1-\lambda^2 \widetilde{T}^2)}{(1-\widetilde{T})^2 (1+\lambda^2 \widetilde{T})}\right].\label{ENonon}
\end{eqnarray}
\end{widetext}

 In the following discussions,
to be more specific, we shall choose two typical values for the channel
transmission $\eta$ in order to study
the entanglement properties after distillation.

\subsection{Pure TMSS: $\eta=1$}
In the literature, PS-based distillation of a pure TMSS has already been numerically treated
 in Ref.~\cite{Kitagawa}. In that work,
 due to the extremely intensive numerical computation
 for diagonalizing a large square matrix, only the low-squeezing regime
$\lambda < 0.9$ was investigated and high photon number terms were neglected.
However, based upon our {\it analytical}
result in Eq.~(\ref{ENonon}), the performance of entanglement distillation
in the large-squeezing (high photon number) regime $0.9<\lambda < 1.0$ can also be examined.

In fact, by simply setting  $\eta=1$, we obtain
\begin{eqnarray}\label{unitetaLogNeg}
&&E_N(\widetilde\rho)_{\eta=1}=\log_2\left[\frac{(1+\lambda)(1-\lambda^2 T)(1+\lambda T)}{(1-\lambda T)(1+\lambda^2 T)(1-\lambda T+\lambda R)}\right],\nonumber\\
&&\\
&& P(on,on)_{\eta=1}=\frac{\lambda^2 (1-T)^2 (1+\lambda^2 T)}{(1-\lambda^2 T)(1-\lambda^2 T^2)}.\label{pe}
\end{eqnarray}

\begin{figure*}
  \includegraphics[width=15 cm]{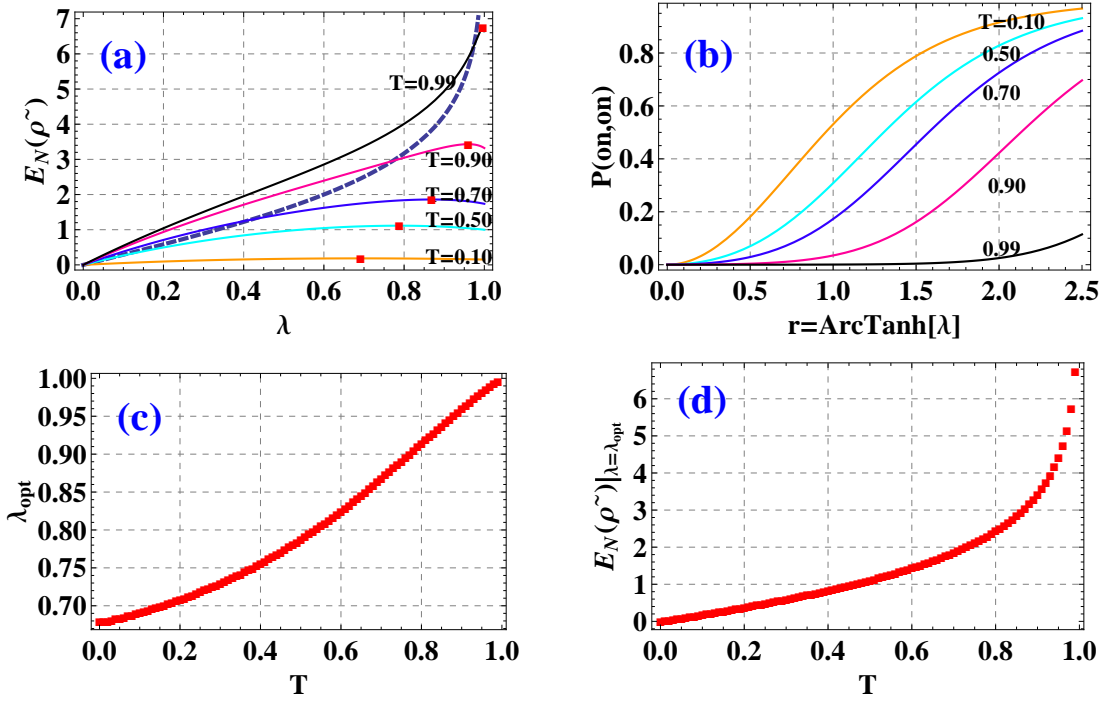}\\
  \caption{(Color online) Performance of distilling a pure TMSS ($\eta=1$) with beamsplitters and on-off detectors. (a) logarithmic negativity of the output state for $T=0.10,0.50,0.70,0.90,0.99$, respectively.
The dashed line corresponds to the logarithmic negativity of the TMSS before distillation (Eq.(\ref{EnrhoBeforeDis}) with $\eta=1$). The red squares in each curve indicate the maximum values of $E_N(\tilde{\rho})$. (b) success probability, i.e., the probability that both detectors record the ``on" results (Eq.(\ref{pe})). (c) $\lambda_{opt}$ as a function of $T$ (see text for more information). (d) Maximal value of $E_N{(\tilde{\rho})}$ at $\lambda=\lambda_{opt}$. }\label{figONOFF}
\end{figure*}

Surprisingly, for a given beamsplitter with finite transmission coefficient $0<T<1$,
 the output entanglement exhibits non-monotonic dependence of the initial
squeezing parameter $\lambda$. The finite transmission
coefficient of the beam splitter has a degrading effect on the output entanglement.
 When $\lambda\rightarrow 1$, a pure TMSS has infinite entanglement.
However, when one uses the beamsplitter together with on-off detectors to implement
the distillation, one will always get finite entanglement.
In fact, the optimal squeezing parameter $\lambda$
(referred to as $\lambda_{opt}$) which maximizes $E_{N}(\widetilde\rho)$ is strictly
 smaller than $1$.
This result is certainly of experimental significance
in order to optimize the distilled entanglement:
 it may not be necessary to prepare as much initial squeezing as possible to maximize the final
entanglement; some finite-squeezing value will be optimal.

In Fig.~\ref{figONOFF} (a), we show the logarithmic
 negativity of the distilled TMSS ($\eta=1$) for different beamsplitter transmissions
 ($T=0.1,~0.5,\cdots,~0.99$). In Fig.~\ref{figONOFF} (b),
we give a plot of the probability of successful distillation, $P(on,on)$.
 Figure~\ref{figONOFF} (c) shows the optimal $\lambda_{opt}$
as a function of $T$, while Fig.~\ref{figONOFF} (d) presents the maximal
 $E_{N}(\widetilde\rho)$ at $\lambda=\lambda_{opt}$. Even with infinite squeezing
 and a non-lossy channel, $\eta=1$, we cannot approach infinite entanglement
after distillation. In fact, when $\lambda\rightarrow 1,$ in Eq.~(\ref{unitetaLogNeg}),
 the logarithmic negativity scales as $E_N(\widetilde{\rho})=\log_2\frac{1}{1-T}$.

In the above distillation protocol, there exists a non-trivial lower bound
 $T_L$ for the transmission coefficient $T$ below which the PS scheme based on
beamsplitters and on-off detectors ceases to improve the entanglement. In Fig.~\ref{figONOFF} (a), it is
 shown that the distillation protocol effectively no longer works for
 $T=0.10,0.50$. The entanglement after distillation is then even smaller than that
 before distillation. Indeed, requiring
 $E_N(\tilde{\rho})>E_N(\rho_{mix})$, the transmission coefficient $T$ satisfies
 $T_L<T \le 1$, with
 \begin{eqnarray}
T_L&=&\frac{1}{3\lambda^3}\left[\lambda(\lambda^2+\lambda-1)+2\sqrt{\xi}\sin\left(\frac{\pi}{6}-\frac{\tilde\theta}{3}\right)\right],\label{TTLL}\nonumber\\
\xi&=&\lambda^2(\lambda^4+2\lambda^3-4\lambda^2+4\lambda+1),\nonumber\\
\zeta&=&\lambda(\lambda^3+8\lambda^2-3\lambda+2),\nonumber\\
\tilde\theta&=&\arccos\left(\frac{3\lambda^3 \zeta-2\lambda(\lambda^2+\lambda-1)\xi}{2\xi\sqrt{\xi}}\right).
\end{eqnarray}
\noindent The quantity $T_L$ in Eq.~(\ref{TTLL}) is a monotonically increasing function of the
squeezing parameter $\lambda$. When $\lambda\rightarrow 0$,
we have $T_L=1/2$. In the other extreme case,
when $\lambda$ approaches $1$, it follows that $T_L\rightarrow 1$.
This, on the other hand, proves the degrading effect of the transmission coefficient $T$:
in the high photon number regime (especially, for $\lambda\rightarrow 1$),
 any finite transmission $0<T<1$ is smaller than $T_L=1$ and the entanglement
 of the state after distillation, as illustrated in Fig.~\ref{SchemefigONOFF},
 is finite and hence smaller than the infinite entanglement before distillation.
 We give a detailed description of the behavior of $T_L$ in Fig.~\ref{TL}.

\subsection{3dB amplitude-damped TMSS: $\eta=0.5$}

\begin{figure*}
  \includegraphics[width=15 cm]{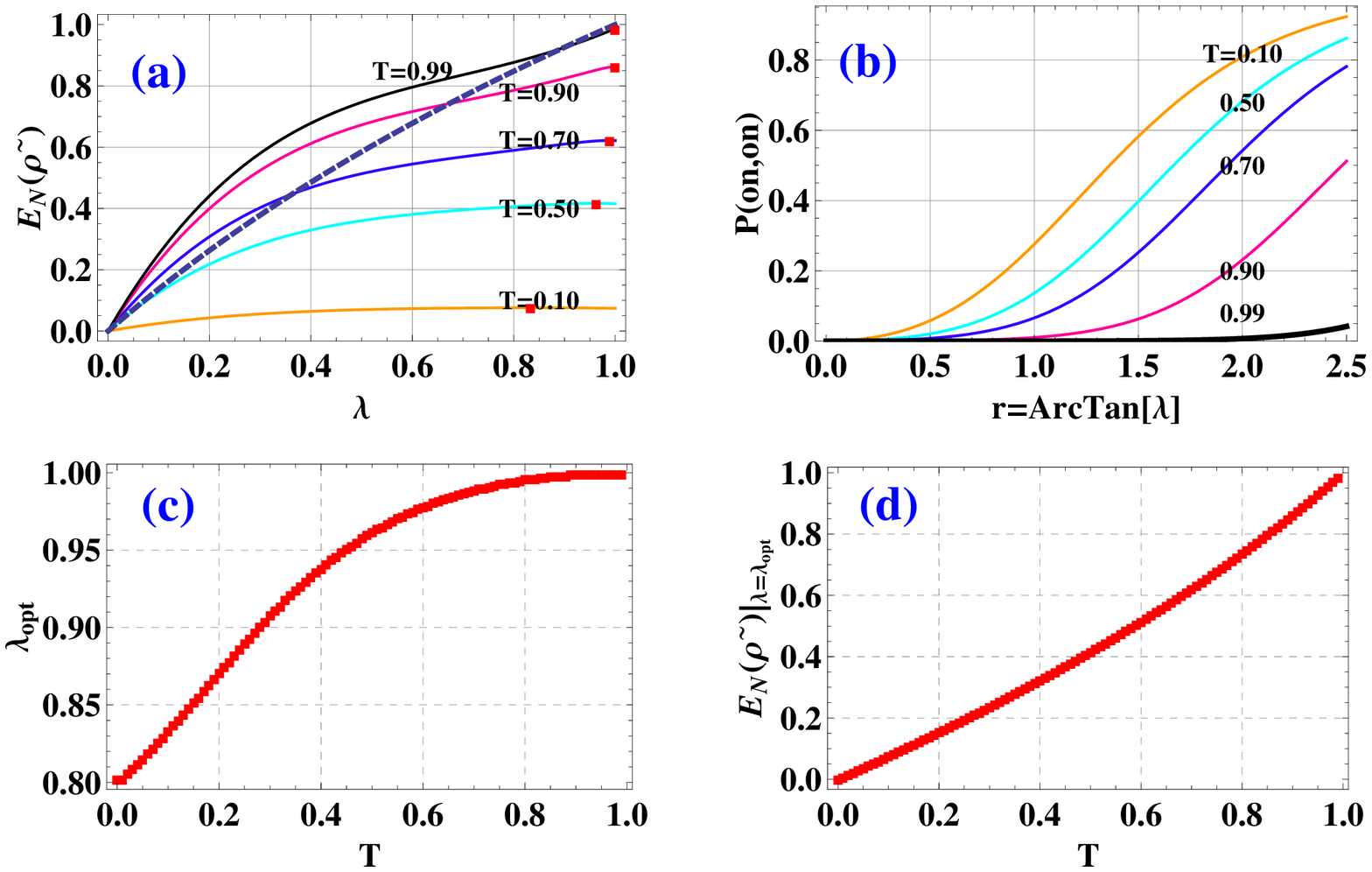}\\
  \caption{(Color online) Performance of distilling a 3dB amplitude-damped
 TMSS ($\eta=1/2$) with beamsplitters and on-off detectors.
 (a) logarithmic negativity of the output state for $T=0.10,0.50,0.70,0.90,0.99$,
 respectively. The dashed line corresponds to the logarithmic negativity of
 the amplitude-damped TMSS before distillation (Eq.(\ref{EnrhoBeforeDis})).
 The red squares indicate the maximum values of $E_N(\tilde{\rho})$.
 (b) success probability of distillation for various $T$ (Eq.(\ref{pe})).
 (c) $\lambda_{opt}$ as a function of $T$. (d) Maximal value of $E_N{(\tilde{\rho})}$
 at $\lambda=\lambda_{opt}$.}\label{figONOFFT0p5}
\end{figure*}

\begin{figure}
  \includegraphics[width=8 cm]{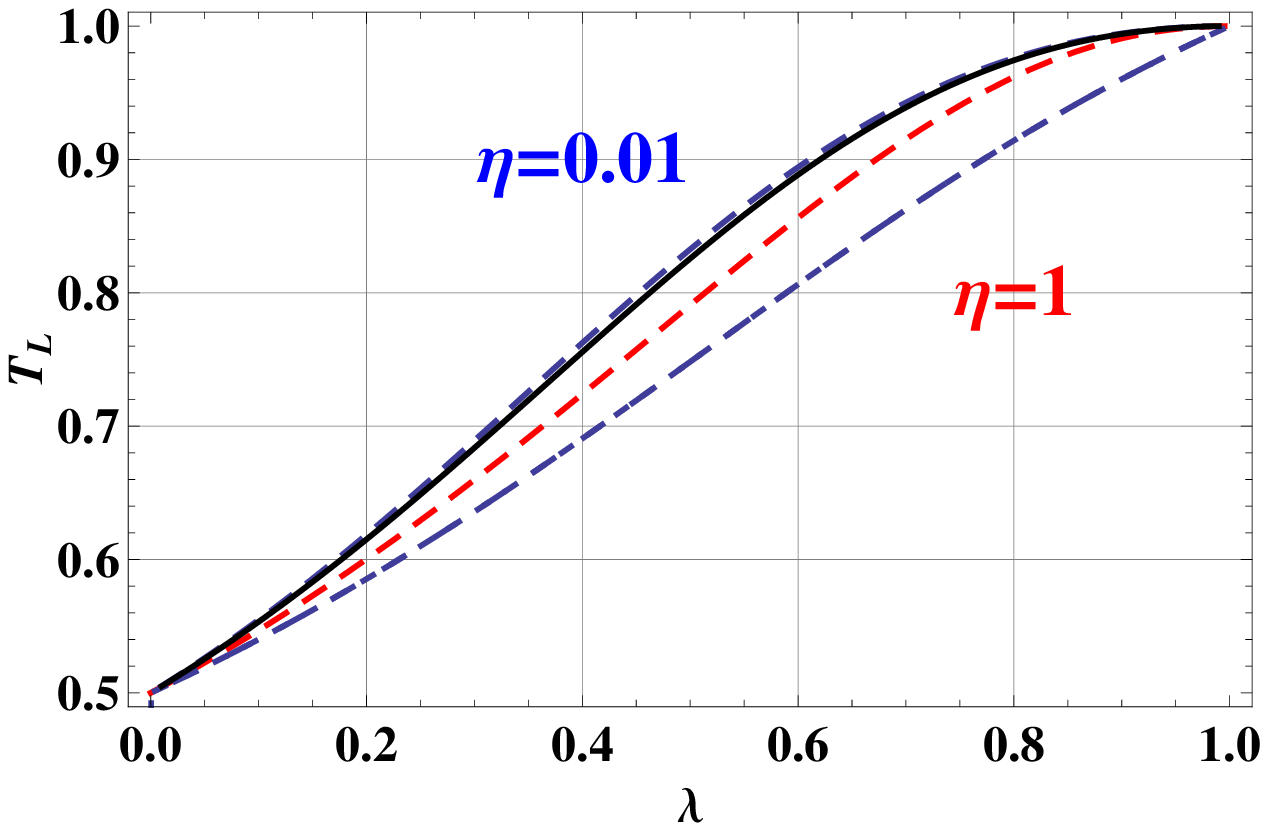}\\
  \caption{(Color online) Lower bound $T_L$ for distilling
 an amplitude-damped TMSS using beamsplitters and on-off detectors.
 The channel transmissions $\eta$ shown, from top to bottom,
 are $0.01,0.1,0.5,1$. $T_L$ increases monotonically with squeezing $\lambda$. }\label{TL}
\end{figure}

In Fig.~\ref{figONOFFT0p5}, we show the success probability and the logarithmic negativity
 of the amplitude-damped TMSS. Here, the amplitude-damping process is simulated
by a 3dB beamsplitter: $\eta=1/2$.
Similar to the distillation of a pure TMSS, the entanglement of the distilled state
 including amplitude damping is again degraded by the finite
 transmission coefficient. There also exists a finite $\lambda_{opt}$ with
$0<\lambda_{opt}<1$ which maximizes the output entanglement. At the same time, the
 lower bound $T_L$ for the transmission coefficient still increases
 monotonically from $T_L=1/2$ to $T_L=1$, when $\lambda$ varies from $0$ to $1$.

In Fig.~\ref{TL}, we show a plot to describe the relation between $T_L$ and
 $\lambda$, for $\eta$ varying from $0.01$ to $1$. It is shown that
for larger channel losses (smaller $\eta$), more transmissive
 beamsplitters (larger $T$) are needed in order to achieve distillation. Furthermore,
 for all $0<\eta\le 1$, the values $T_L$ vary
 from $1/2$ to 1, which means a beamsplitter transmission $T>1/2$ is a general, {\it necessary
condition for distilling amplitude-damped TMSSs using on-off detectors.}
However, in Sec.~\ref{PNRsec}, we will find that such a necessary
 condition can be circumvented
by employing a more sophisticated detection strategy, for instance, using photon number
resolving detectors.

\section{Distillation Using Photon Number Resolving detection}\label{PNRsec}

In quantum communication and computation, using photon number resolving
detectors may lead to various important applications, such as linear-optics quantum computing \cite{KLM},
quantum repeaters \cite{PNRrepeater}, quantum state discrimination \cite{PNRdiscrimi}, and
quantum superresolution \cite{PNRresolution}.
Recently, a photon number resolution of up to $10$
 photons was demonstrated \cite{Afek2009}. In the following,
we shall continue investigating PS-based entanglement distillation protocols, but
we will replace the on-off detectors by PNR detectors.
In our analysis, we will refer to two strategies: (1) pure PNR detection
strategy and (2) mixed PNR detection strategy.

\subsection{Strategy 1: pure PNR detection}

For simplicity, let us consider a perfect PNR detector which has a unique response
for every input photon number state. The corresponding POVM operator for
detecting $\ell$ photons is
\begin{eqnarray}
\hat{\Pi}_\ell=|\ell\rangle\langle \ell|, ~~~~
~~~\sum\limits_{\ell}^\infty\hat{\Pi}_\ell=\mathds{1}.
\end{eqnarray}
This kind of measurement is pure in the sense that the operators $\hat\Pi_\ell$
($\ell=0,\cdots,\infty$) are extremal
in the convex set of all POVMs.
Now suppose both PNR detectors in Fig.~\ref{SchemefigONOFF} give the same photon number $\ell$, then,
according to Eq.~(\ref{rhoABCD}), the output state
 can be written as
\begin{eqnarray}
\widetilde\rho(\ell,\ell)&&=\frac{\mathrm{Tr}_{_{CD}}\left[\rho_{_{ABCD}}\mathds{1}_{_{AB}}\otimes\hat{\Pi}_{\ell_C}\otimes\hat{\Pi}_{\ell_D}\right]}{P(\ell,\ell)}.
\end{eqnarray}
After direct calculation, we obtain the matrix elements of the partially transposed matrix $\rho_{AB}^{\Gamma_{_A}}$ (unnormalized) in
the $K$-subspace,
\begin{widetext}
\begin{eqnarray}
C_{i,j}^{(K)}(\ell,\ell)&&=(1-\lambda^2)\left(\frac{\eta T}{1-\eta}\right)^K\sum\limits_{n=n_0}^\infty(\lambda\eta R)^{2\ell}
(\lambda-\lambda\eta)^{i+j+2n}\sqrt{{i+n+\ell \choose n}{j+n+\ell\choose n} {i+n+\ell \choose i+j+n-k}{j+n+\ell \choose i+j+n-K}}\nonumber\\
&&\times\sqrt{{j+\ell \choose \ell}{i+\ell \choose \ell}{K-i+\ell\choose \ell}{K-j+\ell\choose \ell}},\nonumber\\
n_0&&=\max\{0,K-i-j\},\label{CKdefLL}
\end{eqnarray}
\end{widetext}
as well as the success probability,
\begin{eqnarray}
&&P(\ell,\ell)=\frac{1-\lambda^2}{1-\lambda^2\widetilde T^2}\left[\frac{\lambda \eta R}{1-\lambda^2\widetilde T^2}\right]^{2\ell}\sum\limits_{k=0}^{\ell}{\ell
\choose k}^2(\lambda \widetilde T)^{2k}.\nonumber\\\label{pLL}
\end{eqnarray}
The logarithmic negativity of the output state then becomes
\begin{eqnarray}
&&E_N(\widetilde{\rho}(\ell,\ell))=(2\ell+1)\log_2\left[\frac{1+\lambda\widetilde T}{1-\lambda(\eta T+\eta-1)}\right]\nonumber\\
&&~~~~~~~~~~+\log_2\left[\sum_{k=0}^\ell {\ell\choose k}^2 (\lambda -\lambda\eta)^{2k}(1-\lambda \eta T)^{2\ell-2k}\right]\nonumber\\
&&~~~~~~~~~~-\log_2\left[\sum_{k=0}^{\ell}{\ell\choose k}^2 (\lambda\widetilde T)^{2k}\right].
\end{eqnarray}

In Fig.~\ref{ResLL}, we show the logarithmic negativity and the success probability
 for distilling a 3dB amplitude-damped ($\eta=1/2$) TMSS.
The counted photon numbers are $\ell=1,2,3,4$.
Compared with the distillation using on-off detectors, the PNR-based distillation
 has the following characteristics:
\begin{figure}
  \includegraphics[width=8.5 cm,height=10 cm]{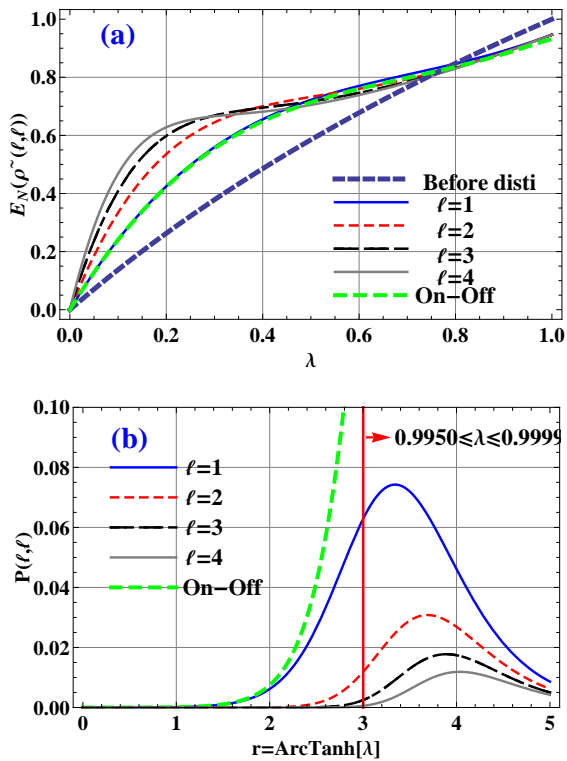}\\
  \caption{(Color online) Comparison of the performance of entanglement distillation
 between PNR detectors (strategy 1) and on-off detectors. The green dashed
 lines correspond to the case of on-off detectors. The blue thick dashed
 lines in (a) represent the entanglement before distillation. Logarithmic negativity
 and success probability are shown for PNR detectors with counted photon numbers
 $\ell=1,2,3,4$. The other parameters are set to $\eta=1/2,T=0.95$.
 The red arrow in (b) indicates the regime
 $0.9950\le\lambda\le 0.9999$ ($3\le r\le 5$).}\label{ResLL}
\end{figure}

\begin{figure}
  \includegraphics[width=8.5 cm]{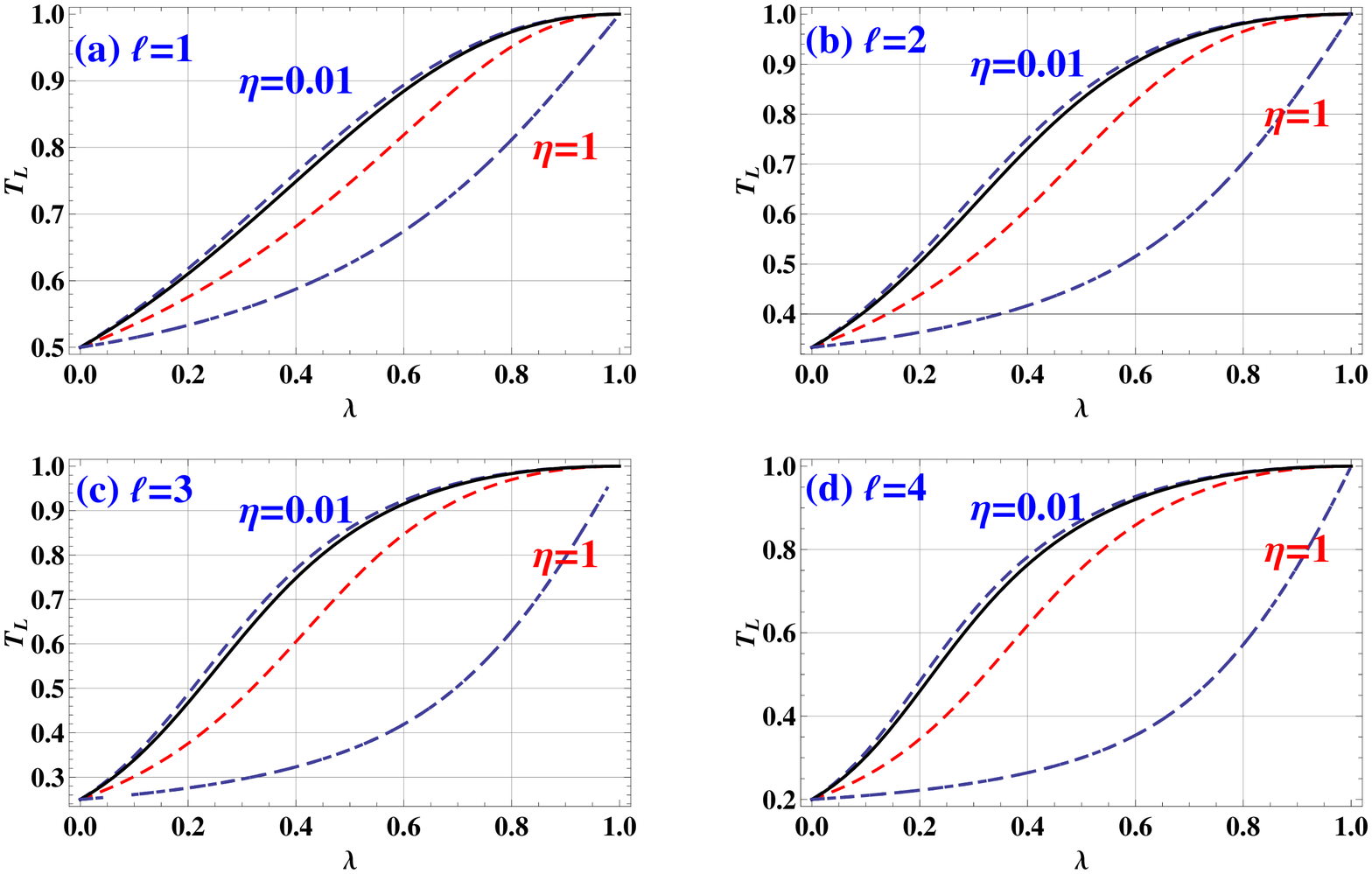}\\
  \caption{(Color online) Lower bound $T_L$ for PNR-based
distillation (strategy 1) of an amplitude-damped TMSS with the numbers
of counted photons $\ell=1,2,3,4$. In each plot (a)(b)(c)(d),
 the channel transmittance $\eta$ is chosen to be $0.01,~0.1,~0.5,~1$,
from top to bottom, and $T_L$ ranges from $1/(\ell+1)$ to $1$.}\label{TLPNR}
\end{figure}

(1) For $\ell\ge 2$, the PNR detectors outperform the on-off detectors
by a significant amount for small squeezing $\lambda$. The more photons are detected,
 the higher the entanglement will be. However, this improvement becomes
 negligible for large squeezing $\lambda$, for which the
 lower bound $T_L$ will be much greater than $T=0.95$
 (the value used in our calculation).

(2) The success probability of the pure PNR distillation strategy decreases exponentially
 with the number of photons detected in each PNR detector, as
can be seen in Eq.~(\ref{pLL}). As a consequence, the probability $P(\ell,\ell)$
is much smaller than the success probability $P(on,on)$ for on-off detectors.
 To be more specific, we show a plot of $P(\ell,\ell)$ as a function of $r={\rm arctanh}(\lambda)$
 in Fig.~\ref{ResLL} (b). In the high-squeezing regime ($0.9950\le\lambda\le 0.9999)$($3\le r
\le 5$), we observe a peak of $P(\ell,\ell)$.
This is because larger squeezing results in more photons in each
 transmitted mode (A and B) and therefore leads to more photons to be
 detected by the PNR detectors. However, too large squeezing will shift
 the number of detected photons to a much higher level $\gg 4$,
 eventually decreasing the detection probability for the $\ell=1,2,3,4$
 photon number cases.

(3) The lower bound $T_L$ for the transmittance of the beamsplitter is
shifted by the PNR detection results. In Fig.~\ref{TLPNR}, we show $T_L$
 as a function of the number of photons detected, $\ell=1,2,3,4$, and the
channel efficiency, $\eta=0.01, 0.1, 0.5, 1$. For $\ell=1$, the bound
 $T_L$ covers the full range between $1/2$ and $1$, similar to
 $T_L$ for the on-off detection protocol (Fig.~\ref{TL}). For
larger $\ell$, e.g. $\ell=2,3,4$, the minimum of $T_L$ (at $\lambda=0$)
is independent of $\eta$ and is shifted to $1/(\ell+1)$, thus
 circumventing the necessary condition $T > 1/2$ for the on-off detection protocols.

\subsection{Strategy 2: mixed PNR detection}

To improve the probability of successful distillation, we introduce another
 distillation measurement
 strategy. This time we shall still use photon number discrimination
 with PNR detectors, however, in a mixed PNR strategy. Such a
 strategy is experimentally more feasible than general
 pure PNR detections and similar experiments have already
 been reported in the context of binary coherent-state discrimination \cite{PNRdiscrimi}.

\begin{figure}
  \includegraphics[width=8.5 cm]{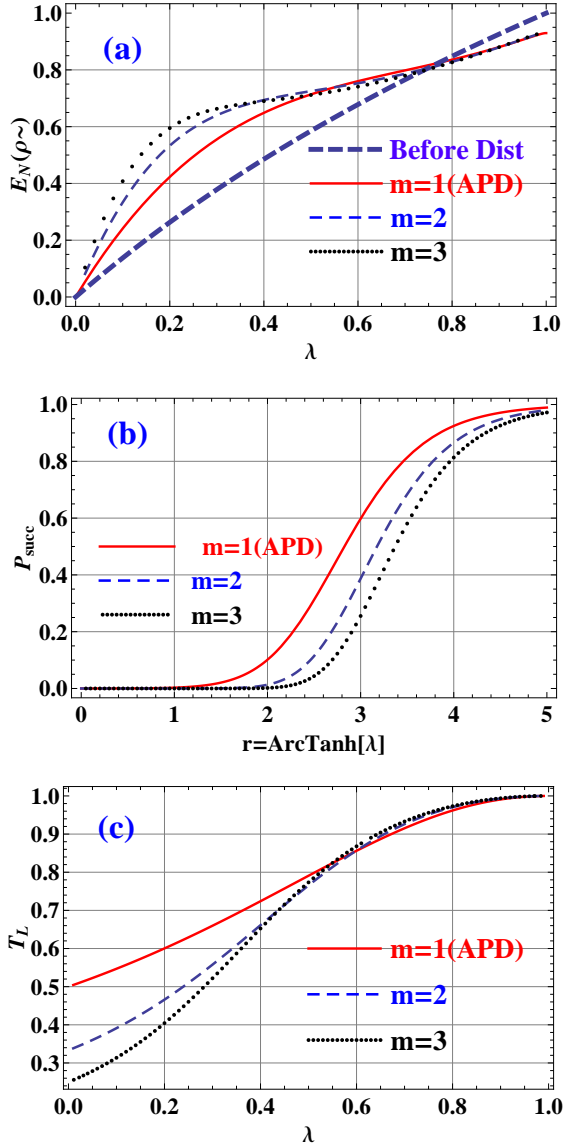}\\
  \caption{(Color online) Logarithmic Negativity (a) and success probability (b) for the mixed PNR
method (strategy 2) with $m=1,m=2,m=3$. The lower bound $T_L$ (c)
varies from $1/(m+1)$ to $1$; the other parameters
are chosen as $\eta=1/2, T=0.95$.}\label{shiftPNR}
\end{figure}

To achieve entanglement distillation, we make a post-selection of the PNR detection
 results and define the POVM operators
\begin{eqnarray}
\hat\Pi_{on}^{(m)}=\sum\limits_{\ell\ge m}^\infty |\ell\rangle\langle \ell|,
~~~~~\hat\Pi_{off}^{(m)}=\sum\limits_{\ell=0}^{m-1}|\ell\rangle\langle \ell|.
\end{eqnarray}
Again, successful distillation is heralded when both PNR detectors
 record the ``on" result. By taking into account the contribution of all
 multi-photon components $\ell\ge m$, the success probability
 approaches $1$ in the case of infinite squeezing ($\lambda\rightarrow 1$).
For any $m$, we have
\begin{eqnarray}
&&P_{succ}^{(m)}=\mathrm{Tr}\left[\rho_{_{ABCD}}
\mathds{1}_{_{AB}}\otimes\hat\Pi_{on_{_C}}^{(m)}\otimes\hat\Pi_{on_{_D}}^{(m)}\right]
\nonumber\\
&=&(1-\lambda^2)\sum_{n=m}^\infty \lambda^{2n}
\left[1-\sum\limits_{k=0}^{m-1}{n\choose k}(\eta R)^{k}\tilde{T}^{(n-k)}\right]^2.\nonumber\\
\end{eqnarray}
When $m=1$, such a strategy is straightforwardly reduced to the
 conventional on-off detection method in Sec.~\ref{on-offSec}.
However, for large $m$, the analytic formulas for success probability $P_{succ}^{(m)}$ and logarithmic
 negativity
$E_N(\widetilde{\rho})$ become rather complicated and we shall only present a
numerical comparison for different $m$ values in Fig.~\ref{shiftPNR}. We still
 consider the typical example of 3dB transmission $\eta=1/2$ and highly transparent
beamsplitters, $T=0.95$.
As can been seen from Fig.~\ref{shiftPNR} (a), for smaller squeezing  $\lambda<0.5$ ($r<0.5493$),
 a significant
increase of entanglement is obtained. For larger squeezing $\lambda>0.5$, the
mixed PNR detection strategy does not improve the entanglement very much.
The corresponding probability $P_{succ}^{(m)}$ is shown in Fig.~\ref{shiftPNR} (b).

Finally, in order to find out for which conditions this mixed-PNR protocol can improve
entanglement, we also systematically vary the $T$ values of the beamsplitters
and calculate the lower bound $T_L$ (Fig.~\ref{shiftPNR} (c)). Interestingly, the $T_L$ bounds
are similar to the pure-PNR case. For $\lambda\rightarrow 0$, a
transmission of $T=1/(m+1)$ is sufficient to enhance the entanglement. However, as $\lambda$
increases, our simulations suggest that a monotonically increasing $T$ is required for
successful distillation.

\section{Operational measure of entanglement}\label{fidelitySec}

In this section, we shall consider quantum teleportation of coherent states
in order to assess the quality of the photon-subtracted entangled states.
Quantum teleportation is a protocol in which an arbitrary, unknown quantum state can be
reliably transferred from a sender to a receiver. The crucial resource for quantum teleportation
to outperform classical teleportation is an entangled
 state shared by the two participants. Intuitively, the more entanglement they share,
 the higher the teleportation fidelity they can achieve. In other words, the
teleportation fidelity may serve as an operational measure of entanglement \cite{Opatrny,Oliver}.

In the following, we consider
a teleportation experiment in which the entangled state after PS-based distillation is
utilized. We assume that the state to be teleported is a coherent state,
 $\sigma_{in}=|\alpha\rangle\langle \alpha|$. Standard continuous-variable teleportation \cite{braunstein98}
consists of three
steps: (1) Alice combines one mode of the entangled state, say the $A$-mode, with the input mode in state $\sigma_{in}$
at a $50:50$ beam splitter; then she measures the quadratures variables $x_{-}=(x_{in}-x_{A})/\sqrt{2}$ and
 $p_{+}=(p_{in}+p_A)/\sqrt{2}$. (2) When she obtains the classical measurement results for $\bar{x}_{-}$ and $\bar{p}_{+}$,
she then communicates them to Bob via a classical communication channel. (3) Using Alice's measurement
results, Bob applies the corrsponding
displacement operation $D(-\beta)=\exp(-\beta a_{_B}^\dagger+\beta^{*} a_{_B}), \beta=\bar{x}_{-}+i \bar{p}_{+}$
on the other entangled mode $B$. The fidelity between $\sigma_{in}$ and the final state of mode $B$
is related with the quality of the shared entanglement. Unit fidelity requires perfect entanglement.

Mathematically, the joint quadrature measurement on the input mode $\sigma_{in}$ and mode $A$ is equivalent to
a heterodyne measurement (acting on mode ${A})$, expressible as \cite{Oliver}
\begin{eqnarray}
\hat{\Pi}_{_A}(\beta)=\frac{1}{\pi}D(\beta)\sigma_{in}^T D^\dagger(\beta),
\end{eqnarray}
where here $T$ denotes the transposition operation. For the normalized entangled state $\widetilde{\rho}_{_{AB}}$,
the probability for outcome $\beta$ is
\begin{eqnarray}
P(\beta)=\mathrm{Tr}[\widetilde{\rho}_{_{AB}}\hat{\Pi}_{_A}(\beta)\otimes \mathds{1}{_{_B}}].
\end{eqnarray}
After the displacement operation by Bob, the final state in mode $B$ becomes
\begin{eqnarray}
 \rho_{{_B}}&&=\frac{1}{P(\beta)}D(-\beta) \mathrm{Tr}_{_A}[\widetilde{\rho}_{_{AB}}\hat{\Pi}_{_A}(\beta)\otimes \mathds{1}_{_B}]D^\dagger(-\beta),
 \nonumber\\
\end{eqnarray}
with a fidelity given by
\begin{eqnarray}
F_{\beta}&&=\langle\alpha|\rho_{_B}|\alpha\rangle\\
&&=\frac{1}{P(\beta)} \langle \alpha+\beta|\mathrm{Tr}_{_A}[\widetilde{\rho}_{_{AB}}\hat{\Pi}_{_A}(\beta)\otimes \mathds{1}_{_B}]|\alpha+\beta\rangle.
\nonumber
\end{eqnarray}

By averaging over all the possible measurement results $\beta$, we obtain the average fidelity
\begin{eqnarray}
F&&=\int d^2 \beta P(\beta) F_{\beta}\nonumber\\
&&=\frac{1}{\pi}\int d^2 \beta \mathrm{Tr}[\widetilde\rho_{_{AB}} D(\beta)\sigma_{in}^T D^\dagger (\beta) \otimes D(\beta) |\alpha\rangle\langle\alpha| D^\dagger (\beta)]\nonumber\\
&&=\mathrm{Tr}[\widetilde\rho_{_{AB}}  \mathbf{O}_\mathrm{F} ],\label{fide}
\end{eqnarray}
where we define the bipartite operator $\mathbf{O}_\mathrm{F}=\frac{1}{\pi}\int d^2 \beta  D(\beta)\otimes D(\beta) (\sigma_{in}^T \otimes |\alpha\rangle\langle\alpha|) D^\dagger(\beta)\otimes D^\dagger(\beta) $.
Using the invariance $d^2 \beta= d^2(\beta+\alpha)$, $\forall\alpha$, and similar methods to those in Ref.~\cite{ladislav}, we find that
\begin{eqnarray}
\mathbf{O}_\mathrm{F}=\sum_{K=0}^\infty\sum_{i,j=0}^\infty\frac{1}{2^{K+1}}\sqrt{{K \choose i}{ K \choose j}} |i,j\rangle\langle K-j,K-i|.\nonumber\\
\end{eqnarray}
Moreover, by noticing that the partially transposed $\mathbf{O}_{\mathrm{F}}^{\Gamma}$ is block diagonal, we can simplify the
fidelity (\ref{fide}) as follows,
\begin{eqnarray}
F=\frac{\mathrm{Tr}[\rho_{_{AB}}^{\Gamma_A}\mathbf{O}_{\mathrm{F}}^{\Gamma}]}{\mathrm{Tr}(\rho_{_{AB}}^{\Gamma_A})}=\frac{\sum_{K=0}^\infty \mathrm{Tr}[ \mathbf{C}_{\mathrm{K}} \mathbf{O}_{\mathrm{F}}^{\Gamma} (\mathrm{K})]}{\sum_{K=0}^\infty \mathrm{Tr}[ \mathbf{C}_{\mathrm{K}}]},
\end{eqnarray}
where $ \mathbf{O}_{\mathrm{F}}^{\Gamma} (\mathrm{K})$ is the $K-$ sub-block matrix
 $ \langle i|\mathbf{O}_{\mathrm{F}}^{\Gamma} (\mathrm{K})|j\rangle = \langle i,K-i|\mathbf{O}_{\mathrm{F}}^{\Gamma}|j,K-j\rangle$.

Thus, using $\mathbf{C}_\mathrm{K}$ as defined above, the teleportation fidelity can be
easily evaluated. For example, for the state before entanglement
distillation, the matrix $\mathbf{C}_\mathrm{K}$ is given by
Eq.~(\ref{CKdef}), and the fidelity becomes
\begin{eqnarray}
F_{mix}&&=\sum_{K=0}^\infty \sum_{i,j=0}^\infty C_{i,j}^{(K)}\cdot \langle i|\mathbf{O}_\mathrm{F}|j\rangle\\
&&=\frac{(1+\lambda)(2-\lambda^3\eta^3+\lambda^2 \eta^2 (\lambda+3)-\lambda\eta(\lambda+4))}
{2(2-2\lambda\eta-\lambda^2\eta+\lambda^2\eta^2)(1-\lambda\eta)(1+\lambda-\lambda\eta)}\nonumber.
\end{eqnarray}

\begin{figure}
  \includegraphics[width=8.5 cm]{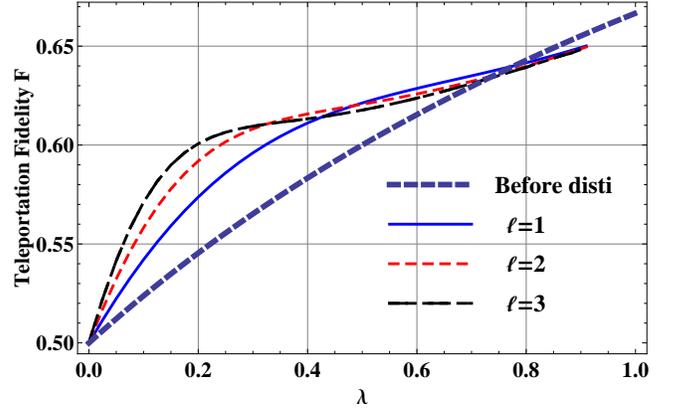}\\
  \caption{(Color online)  Fidelity of teleporting an unknown coherent state
$\sigma_{in}=|\alpha\rangle\langle\alpha|$ using a pure-PNR-distilled
amplitude-damped TMSS. The parameters
are $\eta=1/2, T=0.95$ as in Fig.~\ref{ResLL}.}\label{TeleFidelityPNR}
\end{figure}

Similarly, from the definitions in Eq.~(\ref{CKdefonon}) and Eq.~(\ref{CKdefLL}), the
teleportation fidelity for the PS-distilled states can be obtained, respectively.
For example, in comparison with the
logarithmic negativities calculated in Sec.~\ref{PNRsec}, we
present a numerical evaluation of the teleportation fidelity for pure PNR-distilled entangled states
in Fig.~\ref{TeleFidelityPNR}. For $\ell = 1,2,3$, the teleportation fidelity is obviously improved
in the low-squeezing regime ($\lambda\apprle 0.75$), in a similar way
to what we obtained for the logarithmic-negativity measured entanglement in Fig.~\ref{ResLL}.
However, note that the logarithmic negativity is known to have an operational meaning
(quantified by the quality of quantum correlations used in quantum teleportation)
only for symmetric Gaussian states.
Indeed, our amplitude-damped TMSSs do belong to the class of symmetric Gaussian states.
However, for the photon-subtracted, non-Gaussian states after distillation,
the correspondence between logarithmic negativity and coherent-state teleportation fidelity is not obvious;
even though it is possible to relate the second-moment correlations of photon-subtracted
states with their logarithmic negativities \cite{adessoPSS}.

\bigskip

\section{conclusions}

In conclusion, we have studied a photon-subtraction-based entanglement distillation scheme
on a single copy of a Gaussian mixed state in form of an amplitude-damped two-mode squeezed state (TMSS)
using beamsplitters and various photon detection strategies.
The photon measurements included on-off and
photon number resolving detectors, as well as mixed photon number resolving
detections where the on-off threshold can be varied compared to
the conventional on-off measurement with zero or non-zero photons detected.
Exploiting the symmetry and centrosymmetry properties of the partially transposed
density matrix written in the Fock basis,
we were able to derive explicit formulas for the entanglement of the non-Gaussian mixed states
after distillation in terms of the logarithmic negativity.

We showed that in order to improve the entanglement after the imperfect channel transmission
of the TMSS subject to photon losses, a constraint represented by a lower bound for
the beamsplitters (used for photon subtraction) must be satisfied.
Our results extend earlier work on continuous-variable distillation from pure entangled states to
the more general case of mixed entangled states, as one usually encounters in most
realistic situations such as experimental demonstrations \cite{NatPhotonic}
and optical-fiber-based communications. Most importantly, even for channel attenuations
as large as 3dB, the photon-subtraction-based entanglement distillation scheme
still works fairly well, provided the input squeezing is chosen sufficiently small.

The applicability of our protocol to actual long-distance quantum communication,
for instance, by building up a quantum repeater \cite{duerandbriegel},
depends on various parameters.
First of all, note that the success probabilities in the present scheme
are rather low; i.e., as low as or even lower than those of the known discrete-variable
repeater proposals based upon single-photon detections.
Moreover, our results show that for any (sufficiently small) initial squeezing $\lambda$
for which the distilled entanglement exceeds the input entanglement,
there is always a different, effective squeezing value $\lambda_{\rm eff}>\lambda$
for which the same or even higher entanglement can be distributed through the
lossy channels without subsequent distillations.
This suggests that our distillation still mainly functions as an entanglement concentration,
similar to what can be obtained for photon-subtraction-based distillation
of pure states. It is important to see that distillation still works
for mixed states, however, in a potential application,
it may still be better to use large squeezing from the beginning
without distillation. In this case, the question arises how large this input squeezing
must be to beat the distillation-based protocol.

More specifically, using our formulas, one can find that the logarithmic
negativities before and after distillation are related by
$\lim_{\lambda\to 1}E_N^{\rm before}(\eta,\lambda) >
E_N^{\rm after}(\eta,\lambda_0,T)$, for all initial squeezings
$\lambda_0$, all channel transmissions $\eta$,
and all photon-subtraction transmittances $T<1$.
Nonetheless, for example, with 3dB losses in the channel
(corresponding to an elementary distance in a quantum repeater of almost one attenuation
length), the same entanglement as for transmitting
an almost 10dB-squeezed TMSS without distillation can be obtained through
photon-subtraction-based distillation of a roughly 6dB-squeezed TMSS after transmission.
However, the former approach would be deterministic, whereas the latter
is highly probabilistic, leading to further complications
in a full quantum repeater such as the need for sufficient quantum memories.
Further extensions of our scheme, including more general measurements
and local operations on a single Gaussian mixed state or multi-copy distillations
may prove superior to the protocol presented here.

\section{Acknowledgments}
Support from the Emmy Noether Program of the Deutsche Forschungsgemeinschaft
 is gratefully acknowledged.
SZ acknowledges the support by Max-Planck-Gesellschaft, Chinese Academy of
 Sciences Joint Doctoral Promotion Programme (MPG-CAS-DPP) and Key Lab of Quantum
Information (CAS). The authors thank Jason Hoelscher-Obermaier for discussions.

\end{document}